\documentstyle[preprint,aps,epsf]{revtex}
\tighten
\begin{document}
\draft
\title{Fermion vacuum energies in brane world models}
\author{Antonino Flachi, Ian G. Moss and David J. Toms}
\address{
Department of Physics, University of Newcastle Upon Tyne, NE1 7RU U.K.
}
\date{May 2001}
\maketitle
\begin{abstract}
The fermion representations and boundary conditions in five dimensional anti
de Sitter space are described in detail. In each case the one loop effective
action is calculated for massless fermions. The possibility of topological
or Wilson loop symmetry breaking is discussed. 
\end{abstract} 
\narrowtext

Brane world models contain two types of fields, some restricted to four
dimensional sheets and some living in the higher dimensional bulk
\cite{horova}. The quantum properties of those fields that exist in the
bulk is a rich subject. In this paper we focus on a simple example,
namely massless fermions in five dimensions, and consider the
effective action and the possibility of topological symmetry breaking
\cite{hosotani,ford,dowker1,dowker2}. 

One of the issues which arises is the contribution that vacuum
fluctuations make towards the stability of two parallel branes. This is
important for the Randall-Sundrum scenario, which relates the
mass hierarchy problem to brane separations in anti-de Sitter space
\cite{randall}. The effects of vacuum fluctuations have already been
considered for scalar and fermion fields
\cite{toms,flachi,garriga,hofmann,brevik,goldberger}. However, we feel the need
to clarify some statements that have been made about the fermion boundary
conditions in these models and to calculate the effective action for a 
variety of boundary conditions.

The theory we consider is five dimensional. The fifth dimension is taken to be
an orbifold $S_1/Z_2$, where the circle runs from $y=-a$ to $y=a$ and the $Z_2$
acts by $y\to -y$. The space is equivalent to a five dimensional spacetime
with two four dimensional branes making up the boundary.

The choice of fermion representations in five dimensions is as wide as it is
in four. To start with, we might require that the Lagrangian be invariant
under the full five dimensional Lorentz group. The Lorentz symmetry in this
case would then only be broken by the presence of the brane worlds. The
fermions transform by a matrix $S$, related to a set of gamma
matrices $\Gamma^a$. For the Lorentz transformation $y\to -y$,
\begin{eqnarray} 
S^{-1}\Gamma^5 S&=&-\Gamma^5\\
S^{-1}\Gamma^\mu S&=&\Gamma^\mu\label{trans}
\end{eqnarray}
The smallest representation of the gamma matrices which can satisfy these
relations has eight component spinors. The existence of a matrix $\Gamma^6$
which anticommutes with the other gamma matrices allows $S=i\Gamma^5\Gamma^6$.
The benefits of using eight component spinors have also been emphasised in
discussions of fermion boundary value problems \cite{dowker3}.

The eight dimensional representation can be reduced to a real eight
dimensional Majorana representation, where $\psi=C\psi^*$ and
$C$ satisfies $C^{-1}\Gamma^a C=-\Gamma^{a*}$. These are the fermion
representations which arise from the reduction of supersymmetric theories
\cite{lukas}. In the supersymmetry literature, the eight dimensional fermions
are usually regarded as a pair of four dimensional fermions related by a
symplectic transformation \cite{cremer}. 

If, instead of the full Lorentz group, we require only symmetry under proper
Lorentz transformations, then the representation reduces to four
dimensional Weyl representations. These are the representations that have been
considered hitherto in the context of the Randall-Sundrum scenario
\cite{grossman}. A rule for transforming spinors under the transformation
$y\to -y$ is still required. For this purpose, we can allow the massless Dirac
equation to transform as a pseudoscalar,
\begin{eqnarray} 
S^{-1}\Gamma^5 S&=&\Gamma^5\\
S^{-1}\Gamma^\mu S&=&-\Gamma^\mu
\end{eqnarray} 
The solution is $S=i\Gamma^5=\gamma^5$, where $\Gamma^\mu=\gamma^\mu$ are the
usual gamma matrices in four dimensions. We will refer to these fermions
as `five dimensional Weyl fermions'. They have the useful property that they
induce a chiral particle theory on the branes \cite{grossman}.

The boundary conditions used in this paper are chosen for consistency with
orbifold reductions of the fifth dimension. The fermions carry a
representation of the $Z_2$ symmetry, hence $\psi(y)=\pm S\psi(-y)$. 
Therefore, if 
\begin{equation}
P_\pm=\case1/2(1\pm S),\label{proj}
\end{equation}
we must impose one of the two equivalent conditions $P_\pm\psi=0$ at $y=0$.

The fact that $y$ lies on a circle would normally imply that
$\psi(a)=\psi(-a)$. However, if there is a symmetry it is possible to make the
identification up to a symmetry transformation. The possibility
$\psi(a)=-\psi(a)$ has already been used as a mechanism for breaking
supersymmetry, \cite{scherk}. More general possibilities which would allow
gauge symmetry breaking are discussed later, but for the moment we have two
inequivalent  boundary conditions
\begin{eqnarray} 
y=0:\quad P_-\psi=0,&&\quad y=a:
P_-\psi=0\\ y=0:\quad P_+\psi=0,&&\quad y=a: P_-\psi=0 
\end{eqnarray}
The first set might be regarded as untwisted and the second set twisted. The 
twisted case has been considered in flat five dimensional models by
Antoniadis et al. \cite{antoniadis,a1}. For Weyl fermions, the untwisted
boudary condition agrees with the boundary conditions used by Grossman et
al.\cite{grossman} and gives results similar to Garriga et
al.\cite{garriga}.  We will give results for both Dirac and Weyl fermions and
for both twisted and untwisted cases.

We will take the metric
\begin{equation}
ds^2=e^{-2\sigma}\eta_{\mu\nu} dx^\mu dx^\nu +dy^2.\label{metric}
\end{equation}
Anti de Sitter space corresponds to $\sigma=\kappa y$, $\kappa$ being a
constant. This metric is conformally flat,
\begin{equation}
ds^2=e^{-2\sigma}(\eta_{\mu\nu} dx^\mu dx^\nu +d\tau^2)
\end{equation}
where $0\le\tau\le\beta$, and
\begin{equation}
\beta[\sigma]=\int_0^ae^\sigma dy\label{beta}.
\end{equation}
For anti de Sitter space, $\beta=\kappa^{-1}(e^{\kappa a}-1)$.

When boundaries are present it is convenient to regard the Dirac
operator $D$ mapping one set of fermions into an image set. The adjoint
mapping  is denoted by $D^*$. The one loop contribution
to the effective action is then
\begin{equation} 
W=-\case1/2\log\det(D^*D).
\end{equation}
The boundary conditions on the image fermions can be determined by the
existence of $D^*$. If $P_-\psi=0$, this requires that the normal derivative
of $P_+\psi$ should vanish.

In the massless case, $D=-D^*=i\Gamma^j\nabla_j$ where $j$ runs from $1$ to
$5$. The conformal transformation properties of the massless operator imply 
$D=e^{3\sigma}D_0 e^{-2\sigma}$, where $D_0$ is the Dirac operator in the
strip of flat space $0\le\tau\le\beta$. The boundary conditions are also
conformally invariant. We can relate the effective action to the result in
flat space by 
\begin{equation}
W=W_0+C[\sigma]
\end{equation}
where $W_0=-\case1/2\log\det (D_0^*D_0)$ and $C[\sigma]$ is a
correction term \cite{dowker}. We shall discuss the significance of this term
later.

For untwisted Dirac fields we have $P_-\psi=0$ and 
$\partial(P_+\psi)/\partial\tau=0$ on either boundary.
The eigenvalues of $D_0^*D_0=-\nabla^2$ are then $k^2+m_n^2$, with
$m_n=\pi n/\beta$, where $n=0,1,2\dots$. The degeneracy $g=8$ for each value
of $k$. Twisted Dirac fields have similar eigenvalues with
$m_n=(n+\case1/2)\pi/\beta$. For Weyl fermions, the eigenfunctions and
eigenvalues are unchanged, but now the degeneracy factors are $g=4$ rather
than $g=8$.

The logarithms can be evaluated using $\zeta$-function regularisation
\cite{dowker,hawking}. For untwisted fields,
\begin{equation}
W_0=\int d^4 x\,{3g\over 128\pi^2}\zeta_R(5)\beta^{-4}\label{un}
\end{equation}
where $\zeta_R$ is the Riemann $\zeta$-function. For twisted fields,
\begin{equation}
W_0=-\int d^4 x\,{3g\over 128\pi^2}{15\over 16}\zeta_R(5)\beta^{-4}.
\end{equation}
The results depend on the separation of the branes only through $\beta$ given
in equation (\ref{beta}). The untwisted Weyl case gives the same result as
that obtained by Garriga et. al. \cite{garriga}.

In this particular problem there is no dependence on the
renormalisation scale. The same result can be obtained by dimensional
regularisation where the absence of pole terms also indicates no dependence on
the renormalisation scale. The situation changes when the branes are curved,
and renormalisation scale dependent curvature terms arise \cite{flachi}.

The quantity $W_0$ is also the one loop correction to the effective action of
an infinite set of particles in four dimensions with mass $m_n$. The difference
between $W$ and $W_0$,  namely the cocycle function $C[\sigma]$, can be
regarded as an anomaly in this reduction. Such anomalies have been recognised
by Frolov et al. \cite{frolov,cognola}. In general, $C[\sigma]$ will depend
on the geometry of the branes, but in the present context the anomaly only
contributes a constant term to the matter Lagrangians ${\cal L}_v$ and ${\cal
L}_h$ on the `visible' and `invisible' branes.

In the Randall-Sundrum metric (\ref{metric}), the classical action reduces to
\begin{equation}  
S=-\int d^4x e^{-4\kappa a}\left({\cal L}_v-{3\kappa\over 4\pi G_5}\right)
-\int d^4x \left({\cal L}_h+{3\kappa\over 4\pi G_5}\right),
\end{equation}
where $G_5$ is the gravitational constant in five dimensions. Junction
conditions on the metric imply that both terms vanish in the vacuum. Adding the
correction $W$ for untwisted fermions gives an effective action which now has a
minimum for a particular separation $a$. However, the values obtained cannot
give the correct mass hierarchy ($\kappa a>30$) without a considerable degree
of fine tuning.

We can also derive the vacuum energy density from the effective action quite
simply,  
\begin{equation}
T_{\mu\nu}={2\over \sqrt{g}}{\delta W\over\delta g^{\mu\nu}}
=e^{5\sigma}{dW\over d\beta}g_{\mu\nu}.
\end{equation}
For the untwisted anti-deSitter case, $\sigma=\kappa y$, the energy density is
\begin{equation}
{3g\over 32\pi^2}\zeta_R(5)\kappa^4(e^{\kappa(a-y)}-e^{-\kappa y})^{-5}.
\end{equation}
Since this is strongly concentrated near the visible brane, the back reaction
of this energy modifies the junction conditions, resulting in a value for $a$
in agreement with the minimum of the effective action.

If there is a gauge symmetry, the possibility of topological or Wilson loop
symmetry breaking arises. The boundary conditions can be
generalised by inserting gauge transformations $U_h$ and $U_v$ on the two
branes, so that now
\begin{equation}
P_-=\case1/2(1-SU),
\end{equation}
where $U=U_h$ or $U=U_v$. The condition $P_-^2=P_-$ requires $U_h^2=U_v^2=I$. 
The symmetry is broken, leaving the centraliser of $U_h$ and $U_v$, which
preserves the boundary conditions, as the residual symmetry group.

For a simple, non-trivial, example, consider the group $U(2)$ with
$U_h=\sigma_3$, the Pauli matrix. If $U_v=I$, the residual symmetry group is
$U(1)\times U(1)$ and the fermions decompose into one twisted and one
untwisted fermion. The combined one loop correction is therefore
$\case1/{16}W_0$, where $W_0$ is the untwisted result (\ref{un}). If
$U_v=\pm\sigma_3$, the residual symmetry group is the same but the fermions
are both twisted or both untwisted, giving a correction $2W_0$ or
$-\case{15}/8W_0$. The final case is represented by $U_v=\sigma_1$, and the
residual symmetry group is $U(1)$. The eigenvalues are now of the form 
$k^2+m_n^2$, with $m_n=(n\pm\case1/4)\pi/\beta$. The effective action can be
calculated as before, and takes the value $-\case{15}/{256}W_0$.

For massless fermions, the separation between the two branes is only stable
when the correction to the action is positive. Clearly, there is an interplay
in this scenario between supersymmetry breaking, gauge symmetry breaking and
the mass hierarchy problem. We are presently extending these results to
massive fermions so that we can investigate the consequences in low energy
superstring models.

\acknowledgments

Antonino Flachi is supported by a University of Newcastle Upon Tyne Ridley
Studentship.


\begin{references}
\bibitem{horova}P. Horava and E. Witten, Nucl. Phys.
B460 (1996) 506, Nucl. Phys. B475 (1996) 96 
\bibitem{hosotani}Y. Hosotani, Phys. Lett. B (1983) 309
\bibitem{ford}L. H. Ford, Phys. Rev. D21 (1980) 933
\bibitem{dowker1}J. S. Dowker and S. Jadhav, Phys. Rev. D39 (1989) 1196
\bibitem{dowker2}J. S. Dowker and S. Jadhav, Phys. Rev. D39 (1989) 2368
\bibitem{randall}L. Randall and R. Sundrum, Phys. Rev
Lett 83  (1999) 3370, Phys. Rev Lett 83 (1999) 4670
\bibitem{toms}D. J. Toms, Phys. Letts. B484 (2000) 149
\bibitem{flachi}A. Flachi and D. J. Toms, ``Quantised bulk scalar fields in
the Randall-Sundrum brane model'' hep-th/0103077
\bibitem{garriga}J.Garriga, O. Pujolas and T.
Tanaka, Radion effective potential in the brane world, hep-th/0004109
\bibitem{hofmann}R. Hofmann, P. Kanti and M. Pospelov,
(De)-stabilisation of an extra dimensioan due to a casimir force,
hep-ph/0012213
\bibitem{brevik}I. Brevik, K. A. Milton, S. Nojori and S. D. Odintsov, Quantum
instability of a brane world AdS(5) universe at non-zero temperature,
hep-th/0010205
\bibitem{goldberger}W. D. Goldberger and I. Z. Rothstein, Phys. Lett. B491
(2000) 339
\bibitem{dowker3}J. S. Dowker, J. S. Apps, K. Kirsten and M. Bordag, Class
Quantum Grav. 13 (1996) 2911
\bibitem{lukas}A. Lukas, B. A. Ovrut, K. S. Stelle and D. Waldram,  Phys. Rev
D59 (1999) 086001
\bibitem{cremer}E. Cremer, in `Superspace and supergravity', eds. S. W.
Hawking and M. Ro\~cek (1980) 
\bibitem{grossman}Y. Grossman and N. Neubert, Phys Letts B474 (2000) 361
\bibitem{scherk}J. Scherk and J. H. Schwartz,
Nucl. Phys. B153 (1979) 61 
\bibitem{antoniadis}I. Antoniadis and M. Quiros, Nucl. Phys. B (1997)
\bibitem{a1}I. Antoniadis, S. Dimopoulos and G. Dvali, Nucl. Phys.
B516 (1998) 70
\bibitem{dowker}J. S. Dowker and R. Critchley, Phys. Rev. D16
(1977) 3390 \bibitem{hawking}S. W. Hawking, Commun. Math. Phys. 55 (1977) 133
\bibitem{frolov}V. Frolov, P. Sutton and A. Zelnikov, Phys. Rev. D61
(2000) 024021  
\bibitem{cognola}G. Cognola and S. Zerbini ``On the dimensional reduction
procedure'' hep-th/0008061

\end{references}
\end{document}